\documentclass[aps,showpacs,twocolumn,floats,psfig,superscriptaddress]{revtex4-1}
\usepackage{amssymb}
\usepackage{amsbsy}
\usepackage{amsmath}
\usepackage{amstext}
\usepackage{epsfig}
\usepackage{graphicx, xcolor}  % needed for figures
\usepackage{epstopdf}
\usepackage{units}
	\usepackage{mathtools}
	\usepackage{makeidx}
	\usepackage{amsfonts}
	\usepackage{graphicx, xcolor}  % needed for figures
	\usepackage{dcolumn}   % needed for some tables
	\usepackage{bm}        % for math
	\usepackage{amssymb}   % for math
	\usepackage[ansinew]{inputenc}
	\usepackage[usenames,dvipsnames]{pstricks}
	\usepackage{subfigure}
	\usepackage{epstopdf}
	\usepackage{pst-grad} % For gradients
	\usepackage{pst-plot} % For axes
	\usepackage[colorlinks,hyperindex]{hyperref}
\usepackage[colorlinks,hyperindex]{hyperref}
\usepackage{tikz}
	\hypersetup
	{
		colorlinks,%
		citecolor=black,%
		linkcolor=black,%
		urlcolor=black,%
	}

\newcommand{\be}{\begin{equation}}
\newcommand{\ee}{\end{equation}}
\newcommand{\bea}{\begin{eqnarray}}
\newcommand{\eea}{\end{eqnarray}}
\newcommand{\ba}{\begin{eqnarray}}
\newcommand{\ea}{\end{eqnarray}}

\begin{document}

\title{Dynamics in many-body localized quantum systems without disorder}
\author{Mauro Schiulaz}
\affiliation{SISSA - International School for Advanced Studies, Via Bonomea 265, 34136 Trieste, Italy}
\affiliation{INFN - Sezione di Trieste, Via Bonomea 265, 34136 Trieste, Italy}
\author{Alessandro Silva}
\affiliation{SISSA - International School for Advanced Studies, Via Bonomea 265, 34136 Trieste, Italy}
\affiliation{The Abdus Salam International Center for Theoretical Physics, Strada Costiera 11, 34151 Trieste, Italy}
\author{Markus M\"uller}
\affiliation{The Abdus Salam International Center for Theoretical Physics, Strada Costiera 11, 34151 Trieste, Italy}
\affiliation{Department of Physics, University of Basel, Klingelbergstrasse 82, CH-4056 Basel, Switzerland}

\date{6 May 2015}

\pacs{05.60.Gg 05.30.Rt 64.70.P- 72.20.Ee}
\doi{10.1103/PhysRevB.91.184202}

\begin{abstract}
We study the relaxation dynamics of strongly interacting quantum systems that display  a kind of many-body localization in spite of their translation-invariant Hamiltonian. We show that dynamics starting from a random  initial configuration is nonperturbatively slow in the hopping strength, and potentially genuinely nonergodic in the thermodynamic limit. In finite systems with periodic boundary conditions, density relaxation takes place in two stages, which are separated by a long out-of-equilibrium plateau whose duration diverges exponentially  with the system size. We  estimate the phase boundary of this quantum glass phase, and discuss the role of local resonant configurations. We suggest experimental realizations and methods to observe the discussed nonergodic dynamics.
\end{abstract}

\maketitle

\section{\label{sec:intro}Introduction}

A single quantum particle in a sufficiently strong disorder potential
does not explore the full phase space at given energy, but  remains
Anderson localized in a finite spatial region due to quantum interference~\cite{Anderson}.  Over the last decade it has been shown  that such broken ergodicity and absence
of transport persist in many-body systems of finite density, if disorder
is sufficiently strong  and  interactions are
weak enough~\cite{Anderson-Fleishman-Licciardello,Berkovits-Shklovskii,BAA,Gornyi-Mirlin-Polyakov,Huse-Pal,Oganesyan-Huse, Imbrie}. At non-zero temperature this phenomenon, known as ``many-body localization," comes along with a nonextensive bipartite entanglement entropy in highly excited eigenstates~\cite{Bauer,Pollmann}, and, in well-localized regimes, with a complete set of quasilocal conserved quantities that inhibit transport~\cite{Conserved1,Conserved2,Conserved3,Chandran14}. 

In almost all many-body systems studied so far, quenched disorder is central to stabilization of the localized phase: it ensures
that local rearrangements are typically associated with significant
energy mismatches, which appear as large denominators in perturbation
theory, and suppress real decay processes. In contrast, it was suggested early on in the context of defect diffusion in solid helium crystals~\cite{KaganMaximov} that localization effects could also be induced solely by sufficiently strong interactions, without any quenched disorder. Several recent works have reconsidered this idea, focusing on the question of genuine many-body localization~%\cite{Becca-Fabrizio,Heidelberg, Us,Heidelberg2,Heidelberg3}
in low-dimensional systems, such as Bose-Hubbard models~\cite{Becca-Fabrizio,Heidelberg,Heidelberg2,Heidelberg3},
mixtures of heavy and light interacting particles~\cite{Us} and quantum spin chains~\cite{Nottingham}. In such systems, a tendency to localize arises from the configurational disorder present in generic inhomogeneous initial conditions. 
In Ref.~\cite{Fisher} it was conjectured that another notion of localization, as evinced by an incomplete volume law entanglement, could exist in systems without disorder.

\begin{figure}[h]
\centering
\includegraphics[width=0.45\textwidth, height=0.2\textheight]{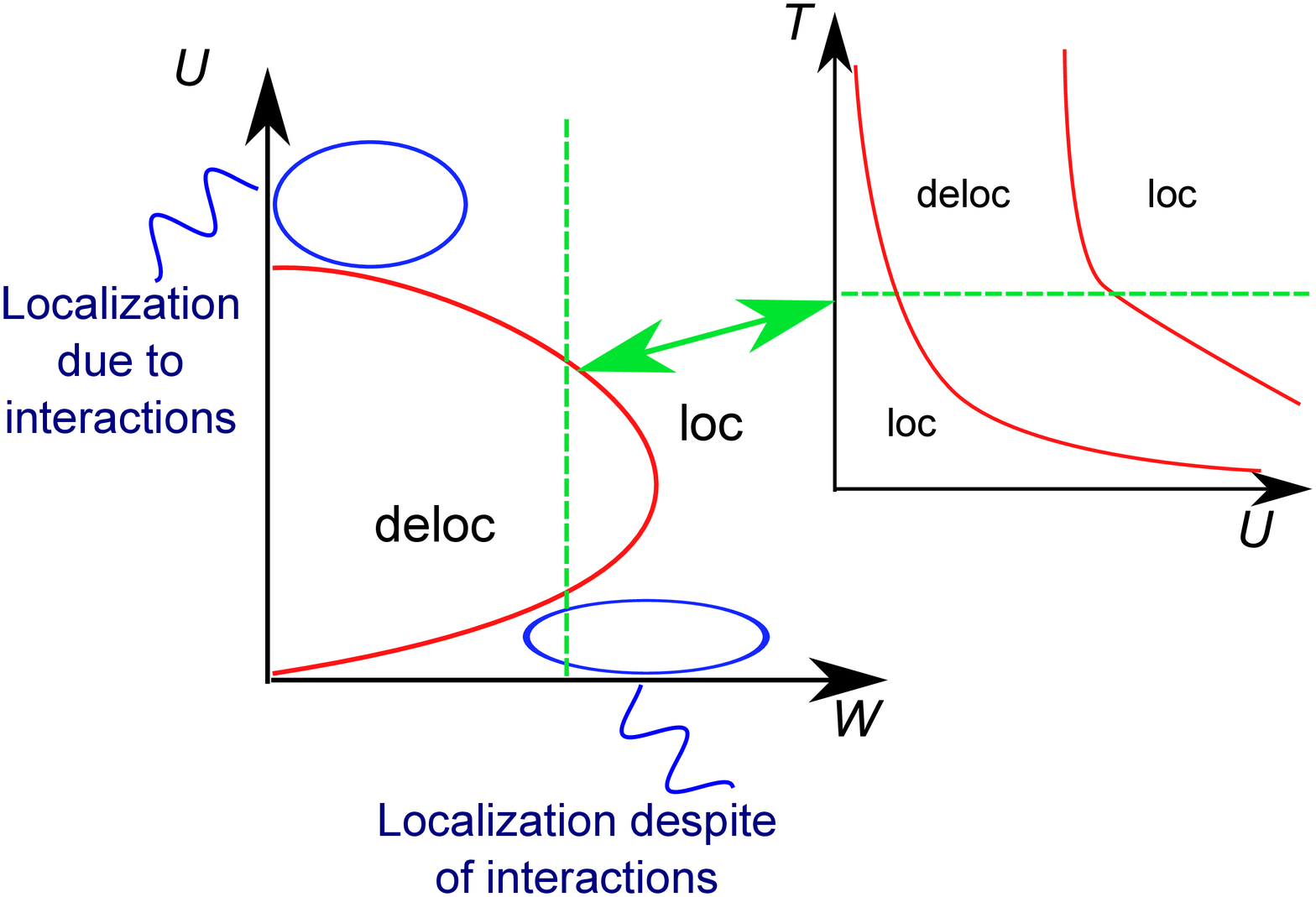}
\caption{{\em Left:} Phase diagram at fixed strength of quantum fluctuations (hopping $t$) and energy density or (quasi-) temperature $T\geq 0$, for one-dimensional models in which commuting interaction ($U$) and disorder ($W$) terms define a classical potential. Both ingredients lead to a rough energy landscape which suppresses quantum tunneling and transport. {\em Right: } The role of temperature differs crucially in the limits of disorder- and interaction-dominated localization: 
%The plot shows a likely scenario for the evolution of the many-body spectrum along the straight line in the main diagram: 
for weak interactions, the lower part of the spectrum is localized, whereas highly excited states are delocalized. The reverse happens when the interaction dominates. 
The dashed lines correspond to a cut at constant quantum fluctuations, disorder, and energy density. They suggest a reentrant localization in the many-body spectrum as interactions are increased.
}
\label{Fig::MBLDiagram}
\end{figure}

Such an interaction-induced localization contrasts in an important way with the more standard scenario~\cite{Anderson-Fleishman-Licciardello,BAA}, in which many-body localization simply embodies the survival of the Anderson-localized phase {\em in spite of dephasing interactions} (see Fig.~\ref{Fig::MBLDiagram}). In the absence of disorder, interactions take a completely different role: they create a rough energy landscape in which weak quantum fluctuations are unable to restore ergodic dynamics, similarly to what happens in classical glasses~\cite{glass1, glass2}. Despite this analogy, in the quantum models of interest to us classical frustration plays no role, in contrast to systems that inherit their nonergodicity from a classically glassy counterpart~\cite{Chamon,Nussinov}. The role of temperature is also opposite to that in disorder-dominated localization, where it enhances the  phase space for scattering and dissipation. When interactions dominate instead, the higher the energy density, the stronger the configurational disorder and hence the localization tendency~\cite{KaganMaximov}.

Localization {\em due to interactions} has the experimentally appealing aspect of being an unambiguous many-body effect, since it cannot be ascribed to disorder. Standard many-body localization manifests itself in the absence of transport and thermalization, but both localized and thermal states are spatially inhomogeneous. In the disorder-free context, however,  the most natural manifestation of localization lies in the dynamical persistence of initial inhomogeneities of particle or energy densities. Such effects are indeed very striking because in any finite system with {\em periodic boundary conditions} and for every inhomogeneous initial condition, translational invariance is eventually restored by the dynamics, as a consequence of momentum conservation. In a localized phase, however, one expects relaxation times to grow exponentially with system size and to diverge in the thermodynamic limit.
Localization effects of this sort could be observed in experiments with binary mixtures of cold atoms~\cite{Gavish04,Gadway10}.
We notice that the idea of detecting localization by the persistence of initial inhomogeneity bears some similarity with the recent, very promising observation of disorder-induced localization through the persistence of an initial density wave pattern,~\cite{Bloch2}. 
 
The remainder of the paper is structured as follows: In Sec.~\ref{sec:model}, we discuss quantitatively the phenomenology of interaction-induced disorder-free localization 
in a class of models of experimental relevance. In Sec.~\ref{sec:inhomogeneity}, we analyze the dynamics in the limit of small quantum fluctuations, and show that, at least within a perturbative treatment, the relaxation of an initial inhomogeneity remains incomplete up to times which are exponentially large in the system size. By extrapolating our result to larger quantum fluctuations, in Sec.~\ref{sec:teff} we will obtain an analytical estimate (upper bound) for the boundary of the localized phase, and in Sec.~\ref{sec:resonances} we estimate the role of local resonant configurations. In Sec.~\ref{sec:bubbles}, caveats related to non-perturbative effects~\cite{Heidelberg2,Heidelberg3}, which might reinstall weak diffusion in very large systems, will be discussed. Finally, in Sec.~\ref{sec:conclusions} we summarize our results, and discuss some experimental setups in which the predicted phenomena might be observed.

\section{\label{sec:model}The model}

\begin{figure}
\centering
\includegraphics[scale=0.4]{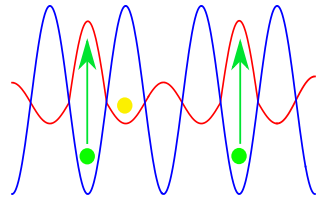}
\caption{{\em Model:} Two atomic species existing on commensurate lattices with different tunneling amplitudes. Heavy particles (green) impede the hopping of light particles (yellow).  Those act as effective springs between the heavy particles, and localize them by creating a complex energy landscape for them.}
\label{Fig::Hamiltonian}
\end{figure}

We consider a quasi-one-dimensional mixture of two interacting hard-core particle species with very different masses~\cite{Us}: a ``fast'' (light) species $a$, and a ``slow'' (heavy) species $c$. As illustrated in Fig.~\ref{Fig::Hamiltonian}, the heavy particles impede the hopping of the light particles. They are therefore referred to as \textit{barriers}. 
The Hamiltonian thus takes the form of an (anti-)assisted hopping model:
\bea
H & = & -J\sum_{j=1}^{L}\left(e^{i\nicefrac{\phi}{L}}a_{j+1}^{\dagger}a_{j}+e^{-i\nicefrac{\phi}{L}}a_{j}^{\dagger}a_{j+1}\right)\left(1-n_{j}\right)\nonumber\\
 & - & t\sum_{j=1}^{L}\left(e^{i\nicefrac{\phi}{L}}c_{j+1}^{\dagger}c_{j}+e^{-i\nicefrac{\phi}{L}}c_{j}^{\dagger}c_{j+1}\right).\label{H}
\eea
The occupation numbers $n_{j}=c_{j}^{\dagger}c_{j}$ are constrained by the conservation of  particles, $\sum_{j}n_{j}=\sum_{j}c^\dagger_{j} c_j = N=\rho L$, and an analogous constraint  for the $a$-particles. The hopping strengths satisfy $ t \ll J$.  We use {periodic boundary conditions} to make the system translationally invariant, but we insert a magnetic flux $\phi$ into the ring so as to break the inversion symmetry. This removes the spectral degeneracy, which simplifies our analysis below. Note that the barriers could equally well be taken to be hard-core
bosons. This choice does not affect the spectrum or localization properties, but only the nonlocal
(in space and time) correlation functions.

The physical essence of this model is retained upon ``integrating out'' the light $a$ particles and substituting them by repulsive springs,  which yields the Hamiltonian
\begin{eqnarray}
H_{\textrm{eff}} & = & -t\sum_{j=1}^{L}\left(e^{i\nicefrac{\phi}{L}}c_{j+1}^{\dagger}c_{j}+e^{-i\nicefrac{\phi}{L}}c_{j}^{\dagger}c_{j+1}\right)+\nonumber \\
 & + & U\sum_{j,l=1}^{L}v\left(l\right)n_{j}n_{j+l}\prod_{k=1}^{l-1}\left(1-n_{j+k}\right),\label{Heff}
\end{eqnarray}
with $v\left(l\right)=l^{-\beta}$. An exponent $\beta=2$ mimics Eq.~(\ref{H}) best at low energies. Indeed,  assume a single fast particle trapped between each pair of successive barriers and assume it to remain in its ground state. The effective repulsion then decays as a power law with exponent $\beta=2$. Note that in this effective model $U$ scales as the hopping, or inverse mass, of the fast particles. Since the phenomenology of disorder-free localization exhibited by the above class of models with an unspecified $\beta>0$ is obviously much more generic than the specific example~(\ref{H}), we focus on the Hamiltonian~(\ref{Heff}) below.

Since our model is translationally invariant, for any finite 
size $L$ the eigenstates can be chosen to be eigenvectors of the discrete translation operator $T$. For infinitesimal hopping $t$, the eigenstates organize in momentum {\em minibands}. These are essentially formed by hybridizations of a classical particle configuration $\left|C\right\rangle$ with all its translations around the ring, $T^{j}\left|C\right\rangle$, for $j=0,1,2,...,L-1$. Typical states correspond to configurations $\left|C\right\rangle$ in which sites are occupied randomly, with probability $\rho$.
The eigenstates of such minibands take the form
\begin{equation}
\left|C,P_n\right\rangle \approx\frac{1}{\sqrt{L}}\sum_{j=0}^{L-1}e^{ijP_n}T^{j}\left|C\right\rangle ,\label{Miniband_state}
\end{equation}
where $P_n$ is the total momentum. The hopping Hamiltonian connects typical configurations $\left|C\right\rangle$ and its translations only at very high order of perturbation theory, since one needs to move all $N=\rho L$ 
particles in order to translate the whole configuration by one site. This leads to an exponentially narrow dispersion of the band
\begin{equation}
\varepsilon_n=-2t_{\textrm{eff}}\cos P_n,\label{Miniband_energy} \quad P_n= (2\pi n+\phi) /L,
\end{equation}
where $t_{\textrm{eff}}$ is the effective hopping of the
center of mass of this state. For small hopping $t$ it is exponentially small in the system size. This is estimated in more detail in Eq.~(\ref{tCM_final}) below. This behavior has important consequences for the dynamics: after preparing the system
 in an inhomogeneous initial configuration, the time scale to relax to a homogeneous state (if averaged over time)
is proportional to $t_{\textrm{eff}}^{-1}$. In the thermodynamic limit, perturbation theory suggests that relaxation is suppressed entirely, and hence the translation symmetry is {\em dynamically} broken.

\begin{figure}
\centering
\includegraphics[scale=0.125]{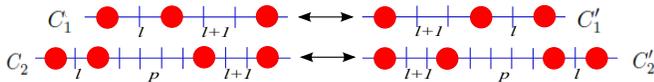}
\caption{Examples of resonances. The configurations $C_1$ and $C_1^{'}$ (top)  hybridize at first order of degenerate perturbation theory in $t$, while $C_{2}$ and $C_{2}^{'}$ (bottom) hybridize at second order. Moving the middle particle(s) to the right costs no energy.}
\label{Fig::Resonance}
\end{figure}

The description of Eq.~(\ref{Miniband_state}) is oversimplified, however, since it neglects the presence of {\em resonances}, i.e., hybridizations with configurations $\left|C'\right\rangle$, which are not translations of $\left|C\right\rangle$, but have the same unperturbed energy.  The simplest types of resonances are shown in Fig.~\ref{Fig::Resonance}: the configurations $C_1, C_1'$  formed by three particles at mutual distances $l$ and $l+1$ are classically degenerate. Their degeneracy is lifted at first order in perturbation theory. In configuration $C_{2}$ the two intervals of lengths $l,l+1$ are separated by an interval of length $p\notin \{l-1,l,l+1\}$, and hence the degeneracy is lifted at second order. In both cases, two hybridized states form:
\begin{equation}
\left|C,P,\pm\right\rangle \approx\frac{1}{\sqrt{L}}\sum_{j=0}^{L-1}e^{ijP}T^{j}\frac{\left|C\right\rangle \pm\left|C^{'}\right\rangle}{\sqrt{2}}.
%\label{hybridization}
\end{equation}
Such states can be seen as the admixture of two of the minibands described
by Eq.~(\ref{Miniband_state}). This can be easily generalized
to the case in which $n$ resonances (labelled $i=1,...,n$) are present in
the initial configuration. Each of them hybridizes a finite number $r_i$ of locally differing, degenerate configurations. The eigenstates then take the form
\begin{equation}
\left|C,P,\left\{ \alpha_{i}\right\} \right\rangle\approx\frac{1}{\sqrt{L}}\sum_{k=0}^{L-1}e^{iPk}
T^{k}
\prod_{i=1}^n \left(\sum_{m_i=1}^{r_i} \psi_{m_{i} }^{ \alpha_{i}} R^{(m_i)}_i \right)
\left|C\right\rangle
\end{equation}
where the $\left\{ \alpha_{i}\right\} $ label the possible states of the $i$'th resonance. Those are described by amplitudes $
\psi^{\alpha_i}_{m_i}$ multiplying local operators $R_i^{(m_i)}$ that rearrange the classical configuration at the resonant spot. 

The restriction to exactly resonant hybridizations applies for very small $t$ only. At larger hopping, states with finite energy differences of $O(t)$ hybridize as well. Nevertheless, the crucial point of the analysis of Ref.~\cite{Us} is that at the perturbative level in $t/U$ no system spanning hybridizations are expected. This is expected despite the fact~\cite{Abanin} that in the thermodynamic limit the exponentially many minibands (\ref{Miniband_energy}) overlap in energy, because the matrix elements between most minibands are even much smaller than the level spacings resulting from band overlaps.

\section{\label{sec:inhomogeneity}Temporal decay of spatial inhomogeneity}

Let us now consider the time evolution from a classical initial configuration $C$. We first restrict consideration to the case where $\left|C\right\rangle $ has no resonant spots, which allows for exact calculations.
To characterize the relaxation process, we define the average spatial density inhomogeneity,
\begin{equation}
\Delta\rho_{\psi}^{2}\left(\tau\right)\equiv\frac{1}{L}\sum_{j=1}^{L}\left[\left\langle \psi\left(\tau\right)\right|\left(n_{j+1}-n_{j}\right)\left|\psi\left(\tau\right)\right\rangle \right]^{2},
\end{equation}
where $\left|\psi\left(\tau\right)\right\rangle \equiv e^{-iH\tau}\left|C\right\rangle $. This 
observable vanishes for any translationally invariant state, and can be measured in cold-atom experiments using microscopy techniques~\cite{Greiner,Bloch,Bloch2}.
Below, we will also consider its time-average, $\left\langle \Delta\rho_{\psi}^{2}\right\rangle(T)\equiv T^{-1}\int_0^T d\tau \Delta\rho_{\psi}^{2}\left(\tau\right) $, which will be insensitive to quantum revivals in finite systems.
In the absence of resonances, the relevant eigenstates and energies are given by Eqs.~(\ref{Miniband_state}) and (\ref{Miniband_energy}), and one finds (see Appendix~\ref{sec:appdecay})
\begin{eqnarray}
\Delta\rho_{\psi}^{2}\left(\tau\right) & = & \frac{1}{L^{4}}\sum_{m\neq n=0}^{L-1}\sum_{n^{\prime}\neq m^{\prime}=0}^{L-1}e^{-i\tau\left[\left(\varepsilon_{n}+\varepsilon_{n^{\prime}}\right)-\left(\varepsilon_{m}+\varepsilon_{m^{\prime}}\right)\right]}\nonumber\\
&\times&\sum_{k=0}^{L-1}\sum_{k^{\prime}=0}^{L-1}e^{i\frac{2\pi}{L}\left(m-n\right)k}e^{i\frac{2\pi}{L}\left(m^{\prime}-n^{\prime}\right)k^{\prime}}\nonumber \\
 & \times & \left[2G\left(k-k^{\prime}\right)-G\left(k-k^{\prime}-1\right)\right.\nonumber\\
& - &\left. G\left(k-k^{\prime}+1\right)\right],\label{DrhoT}
\end{eqnarray}
with the  auto-correlation function of the initial density,
\begin{equation}
G\left(k-k^{\prime}\right)\equiv\frac{1}{L}\sum_{j=1}^L\left\langle C\right|n_{j+k}\left|C\right\rangle \left\langle C\right|n_{j+k^{\prime}}\left|C\right\rangle.
\end{equation}

In the thermodynamic limit, we can take a continuum limit and measure time naturally in units of the inverse of the effective center of mass hopping, $t_{\rm eff}^{-1}$. Assuming an essentially random initial configuration of particles of density $\rho$, we further have $G(k-k^{\prime})=\rho\left(1-\rho\right)\delta_{k-k^{\prime},0}+\rho^{2}$. After some manipulations one finds that the inhomogeneity relaxes according to
\begin{equation}
\frac{\Delta\rho_{\psi}^{2}\left(\tau\right)}{\Delta\rho_{\psi}^{2}\left(0\right)}=\intop_{-\pi}^{\pi}\frac{dq}{2\pi}J_{0}^{2}\left(4\tau t_{\rm eff}\left|\sin q\right|\right)\sin^{2}q,\label{Delta_rho_infinity}
\end{equation}
where $J_{0}$ denotes the Bessel function of the first kind.
For times $\tau \ll t_{\rm eff}^{-1}$ one finds essentially no relaxation,
\begin{equation}
\frac{\Delta\rho_{\psi}^{2}\left(\tau\right)}{\Delta\rho_{\psi}^{2}\left(0\right)}=1-6(\tau t_{\rm eff})^{2}+O\left((\tau t_{\rm eff})^{4}\right),\quad\tau \ll t_{\rm eff},
\end{equation}
reflecting the absence of any local resonances.
For large times, if no time average is taken the inhomogeneity oscillates, with an envelope decaying as $\Delta\rho_{\psi}^{2}\left(\tau\right)\propto\tau^{-1}$.

\begin{figure}
\centering
\includegraphics[width=0.5\textwidth]{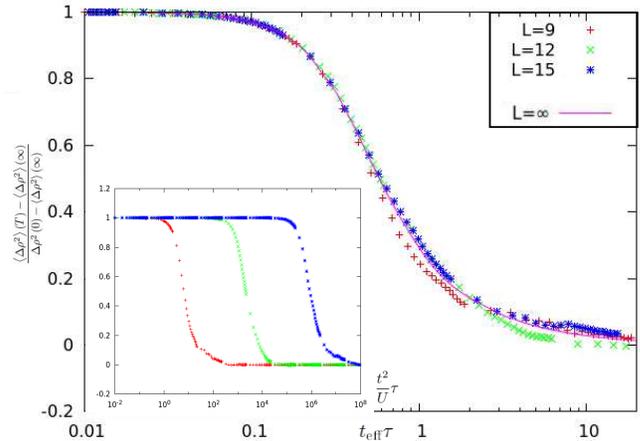}
\caption{Relaxation of  inhomogeneity in the density, in the absence of resonances. Time is rescaled by the exponentially large sample-dependent $t_{\rm{eff}}^{-1}$. The solid line is the analytical result~(\ref{Delta_rho_infinity}) for the thermodynamic limit. {\em Inset:} The same numerical data without rescaled time show that the density inhomogeneity persists for times which diverge with the system size.}
\label{Fig:Drinfinity}
\end{figure}

In Fig.~\ref{Fig:Drinfinity} we compare the above calculations with numerical data from exact
diagonalization of finite systems, initialized in a configuration $C$ of $N=\rho L$ particles, with $\rho=1/3$. {We have restricted the numerics to configurations that do not exhibit resonances at any order in perturbation theory.} We used very small hopping $t=10^{-3}U$ and interactions  decaying with an exponent $\beta=2$. For each data set,  time is rescaled with the appropriate effective center-of-mass-hopping, $t_{\rm eff}(C)$.
In finite systems,  the long time average
of $\left\langle \Delta\rho_{\psi}^{2}\right\rangle\left(T\right)$ is finite, and
for a nondegenerate spectrum a simple calculation yields
$\left\langle \Delta\rho_{\psi}^{2}\right\rangle\left(\infty\right) =\Delta\rho_{\psi}^{2}\left(0\right)/L$. This
is subtracted in Fig.~\ref{Fig:Drinfinity}, so that all curves asymptotically tend to zero. Despite the small sizes, the agreement with Eq.~(\ref{Delta_rho_infinity}) for the thermodynamic limit is very good. 

\section{\label{sec:teff} Estimate of the effective hopping}% $t_{eff}$}

The inset of Fig.~\ref{Fig:Drinfinity} illustrates the long-time plateau of inhomogeneity, whose length diverges exponentially in the thermodynamic limit.
The latter is due to the exponential smallness of $t_{\rm eff}$, $\ln\left(t_{\rm eff}\right)\propto -L$.

Let us estimate that quantity. We first consider a configuration $C$ which exhibits no resonances at any order of perturbation theory. This means that the displacement of any subset of $n<N$ particles by one site (all in the same direction) does not lead to a configuration whose classical energy is degenerate with that of $C$. This restriction is equivalent to requiring that no two intervals between successive particles differ by one lattice spacing only. In this special case we can compute the effective hopping $t_\textrm{eff}$ using ordinary nondegenerate $N$th-order perturbation theory in $t$. We need to sum over all possible orders in which we can move $N$ particles forward by one site each and divide the hopping matrix elements by the corresponding intermediate energies. This leads to the expression
\be
t_\textrm{eff}=t\sum_{P\in S(N)}\prod_{i=1}^{N-1}\frac{t}{\sum_{j=1}^{i}\Delta V_{P\left(j\right),P}^{\textrm{exact}}}. \label{teff_formal}
\ee
$P$ runs over all permutations of $N$ elements, and $\Delta V_{P\left(j\right),P}^{\textrm{exact}}$ is the energy shift associated with the displacement of particle $P\left(j\right)$. It has an explicit dependence on the permutation $P$, as the energy shift depends on whether or not particles $P\left(j\right)\pm1$ have already moved when particle $P(j)$ moves:
\begin{widetext}
 \be
\frac{\Delta V_{P\left(j\right),P}^{\textrm{exact}}}{U}=\begin{cases}
v\left(l_{P\left(j\right)}+1\right)-v\left(l_{P\left(j\right)}\right)+v\left(l_{P\left(j\right)+1}-1\right) -v\left(l_{P\left(j\right)+1}\right) & \textrm{if neither }P\left(j\right)\pm1\textrm{ have moved before  step $j$,}\\
v\left(l_{P\left(j\right)}+1\right)-v\left(l_{P\left(j\right)}\right) +v\left(l_{P\left(j\right)+1}\right) -v\left(l_{P\left(j\right)+1}+1\right)  & \textrm{if only }P\left(j\right)+1\textrm{ has moved  before step $j$,}\\
v\left(l_{P\left(j\right)}\right)-v\left(l_{P\left(j\right)}-1\right)+ v\left(l_{P\left(j\right)+1}-1\right)-v\left(l_{P\left(j\right)+1}\right) & \textrm{if only }P\left(j\right)-1\textrm{ has moved  before step $j$,}\\
v\left(l_{P\left(j\right)}\right)-v\left(l_{P\left(j\right)}-1\right) +v\left(l_{P\left(j\right)+1}\right) -v\left(l_{P\left(j\right)+1}+1\right) & \textrm{if both }P\left(j\right)\pm1\textrm{ have moved before step $j$.}
\end{cases}
 \ee
\end{widetext}
Here $l_{j}\equiv\left|r_{j}-r_{j-1}\right|$ is the distance between particles $j$ and $j-1$. The interaction $v\left(l\right)$ is the one appearing in the effective Hamiltonian~(\ref{Heff}).

Expanding the interaction energies in the distance, we can rewrite this as
\be
\Delta V_{P\left(j\right),P}^{\textrm{exact}}=\Delta V_{P(j)}^{\left(1\right)}  +U\delta V_{P(j),P},\quad \delta V_{P(j),P} =O\left(v''(l)\right),
\ee
where the leading term at low density ($\rho\ll1$ and thus, typically, $l_j\gg 1$),
\be
\Delta V_{i}^{\left(1\right)} =U\left[v'(l_{i})-v'(l_{i+1})\right],
\ee
does not depend on $P$ explicitly.

Let us first discuss the sum over permutations qualitatively. Even though there are $N!$ terms, most of them have denominators that grow factorially as well. Given that the $\Delta V$ have essentially random signs, typical denominator products scale as $\sqrt{N!}$ and have random signs, too. This compensates the factorial number of (randomly signed) terms and leaves us with a merely exponentially growth with $N$.

Next, we observe that for $N$ numbers
$A_{1},A_{2},...,A_{N}$, it holds that
\begin{equation}
\sum_{P\in S(N)}\prod_{i=1}^{N}\frac{1}{\sum_{j=1}^{i}A_{P\left(j\right)}}=\prod_{j=1}^{N}\frac{1}{A_{j}},\label{Permutations_sum}
\end{equation}
which is easily proved by induction.
We can apply this result  to Eq.~(\ref{teff_formal}), taking $A_{j}=\Delta V_{j}^{\left(1\right)}$. This yields
\begin{equation}
t_{\textrm{eff}}\simeq t^{N}\left(\prod_{i=1}^{N-1}\frac{1}{\Delta V_{i}^{\left(1\right)}}\right)\sum_{j=1}^{N}\Delta V_{j}^{\left(1\right)}.\label{tCM}
\end{equation}
The product term suggests an exponential behavior of $t_{\rm eff}$ with $N$, as anticipated above.

However, the prefactor of $N$ in the exponent is not estimated correctly by this calculation. Indeed, we notice that the sum $\sum_{j=1}^{N}\Delta V_{j}^{\left(1\right)}$ vanishes exactly. This implies that terms of higher order in $\rho$ must be retained to obtain a finite result. This is a nontrivial interference effect affecting the motion of clusters of particles.
Analytically it is  difficult to treat such higher order corrections, since they depend on the order in which particles move. 
However, we have obtained lengthy analytical expansions in $\rho$ for small $N\leq 5$, which show that there are $N-1$ extra factors 
 of the form $v''(l)/v'(l)$, which scales as $v''(l)/v'(l) \sim 1/l\sim \rho$ for any power-law interaction.
This indicates that the first nonvanishing term presumably scales as 
\bea
t_{\textrm{eff}}\sim t_{\textrm{eff}}^{(1)}\rho^{N-1}, \quad \rho\ll 1, \label{rhoscaling}
\eea
for all $N$, whereby
\be
t_{\textrm{eff}}^{(1)}= t^{N}\left(\prod_{i=1}^{N-1}\frac{1}{\Delta V_{i}^{\left(1\right)}}\right),
\ee
is the leading result which one naively expects from Eq.~(\ref{tCM}) {and similar estimates in Ref.}~\cite{KaganMaximov}.

We have verified this behavior numerically, by studying the scaling of the exact expression~(\ref{teff_formal}) with the density for small $N$. In Fig.~\ref{Fig::Permutations}  we plot the ratio $\nicefrac{t_{\textrm{eff}}^{\textrm{exact}}}{t_{\textrm{eff}}^{(1)}}$, logarithmically averaged over non-resonant configurations, with an exponential distribution of interval lengths of mean $\rho^{-1}$. The numerical data are indeed consistent with Eq.~(\ref{rhoscaling}).
This leads to the following estimate, valid to logarithmic accuracy:
\begin{equation}
\ln\left(\frac{t_{\textrm{eff}}}{t}\right)_{{\rm typ}}\approx(N-1)\left[\left\langle \ln\left(\frac{t}{\Delta V^{\left(1\right)}}\right)\right\rangle +\ln\rho+c_\beta\right].
\label{tCM_final}
\end{equation}
The angle brackets indicate an average  over all particles and $c_\beta$ is a constant that depends on the exponent $\beta$ characterizing the interactions between particles ($c_2\approx4$).

\begin{widetext}
\begin{center}
\begin{figure}[t]
\includegraphics[scale=0.525]{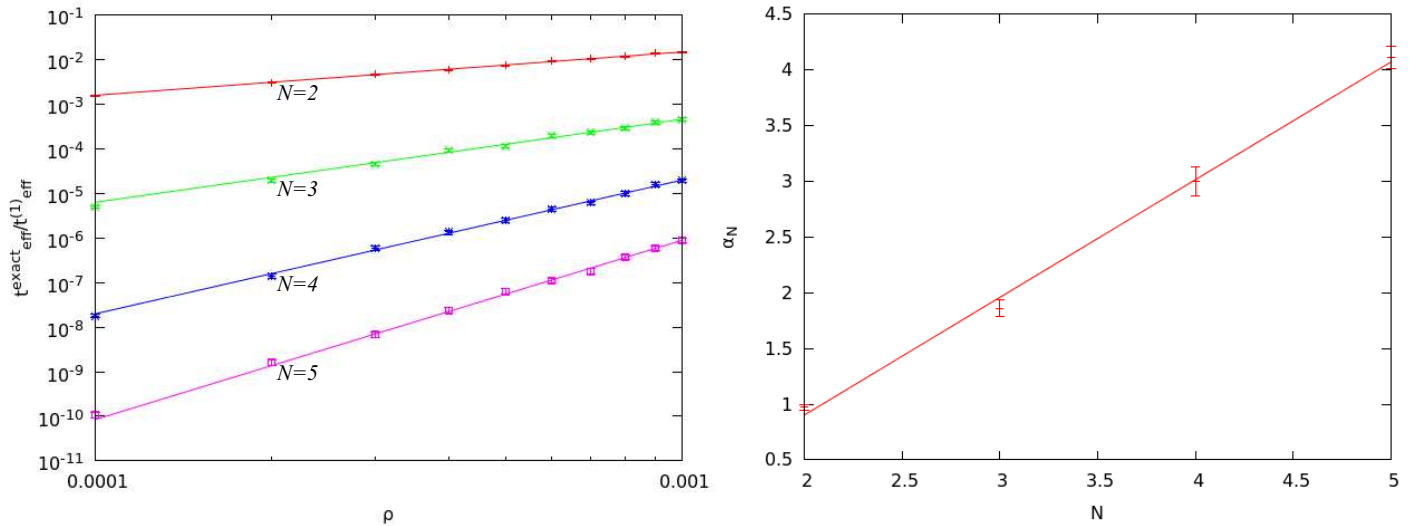}
\caption{{\em Right:} Logarithmically averaged ratio between the exact hopping $t_{\textrm{eff}}^{\textrm{exact}}$ and the naive estimate $t_{\textrm{eff}}^{(1)}$, as a function of density $\rho$, for different numbers of particles. The ratio was found to scale as $\rho^{\alpha_N}$. {\em Left:} A plot of the fitted $\alpha_N$ against $N$  confirms that $\alpha_N =N-1$ (solid line).}
\label{Fig::Permutations}
\end{figure}
\end{center}
\end{widetext}

The above estimates are quantitatively good only for very small $t$. We may nevertheless use them to estimate the hopping $t_c$ at which typical random states delocalize, by requiring that the coefficient of $N$ on the left-hand side of Eq.~(\ref{tCM_final}) vanishes. More precisely, we expect
\begin{equation}
\frac{t_{c}}{U}\lesssim\rho^{-1}  \exp[\left\langle \ln(\Delta V^{\left(1\right)} /U) \right\rangle -c_\beta],
\label{t_C}
\end{equation}
to be an upper bound,  since locally resonating structures proliferate with increasing $t$. 
For the power-law interactions $v\left(l\right)=l^{-\beta}$ considered here, one finds $t_{c}\sim 
U \rho^{\beta}$ to be of the order of the typical interparticle interaction. Due to the many-body interference effect discussed above, this is larger by $\rho^{-1} \sim l_{\rm typ}$ than the naive expectation that $t_c$ should be of the order of typical interaction {\em forces} between particles, as would be predicted by using $t_{\rm eff}^{(1)}$ for this estimate.

To estimate the value of $t_c$ at the moderate density $\rho=1/3$ and for $\beta=2$, we have fitted the size dependence of the numerically evaluated $t_{\rm eff}$ as $t_{\rm eff}\propto(t/t_c)^{N}$ where $N=\rho L$. This yielded 
\bea
t_c\left(\rho=1/3\right)\approx0.2U.
\eea
 This is quite consistent with the numerical results of the recent work~\cite{Abanin}.

\section{\label{sec:resonances}Effect of local resonances}

Let us now discuss the role of local resonances in the configurations $C$.
% that is, compact subsets of the configuration $C$, which, when moved by one lattice spacing to the right, lead to a new configuration $C'$ with the same classical energy as $C$. 
%At low density the dominant type of resonances is shown in Fig.~\ref{Fig::Resonance}. 
It is still expected that at small enough $t$, the effective hopping of the center of mass of a generic configuration $C$ scales as $t_{\textrm{eff}}\propto t^{\alpha N}$. Resonances simply reduce the exponent $\alpha$ with respect to the naive expectation $\alpha=1$. To understand the origin of this effect, let us consider the simple case of three particles on a ring. We call the three interparticle distances $l_1,l_2,l_3$. In general, the effective hopping in this system is proportional to $\nicefrac{t^{3}}{U^{2}}$, since to translate the entire system all particles must be moved by one site. Now let us analyze the resonant case $l_2=l_1+1$. As illustrated in Fig.~\ref{Fig::Resonance}, there are two degenerate configurations, which form the hybridized states
\be
\left|\psi_{\pm}\right\rangle \approx\frac{\left|l_{1},l_{1}+1,l_{3}\right\rangle \pm\left|l_{1}+1,l_{1},l_{3}\right\rangle }{\sqrt{2}}
\ee
with an energy splitting of order $O(t)$.
It is straightforward to see that a matrix element between $\left|\psi_{\pm}\right\rangle$ and the translated wavefunctions $T\left|\psi_{\pm}\right\rangle$ appears already at second order in $t$, not only at third order. This implies that the effective hopping of this configuration is only of order $\nicefrac{t^{2}}{U}$.

An alternative way of understanding this result is as follows. If two resonant intervals are present, the ensuing degeneracy of the spectrum is split at first order  in perturbation theory if the intervals are direct neighbors. If they are not adjacent to each other and if they are surrounded by intervals of different lengths, the splitting is generically of second order $\sim\nicefrac{t^2}{U}$. %This provides them with an energy splitting of order $t$, or $\nicefrac{t^2}{U}$, respectively. 
In the calculation of the effective hopping, such lifted resonances appear as small denominators, which increase the transition amplitude by one or two factors of $\nicefrac{U}{t}$, respectively. This argument is easily generalized to configurations with multiple, spatially distant resonances.

\begin{figure}[h]
\centering
\includegraphics[scale=0.5]{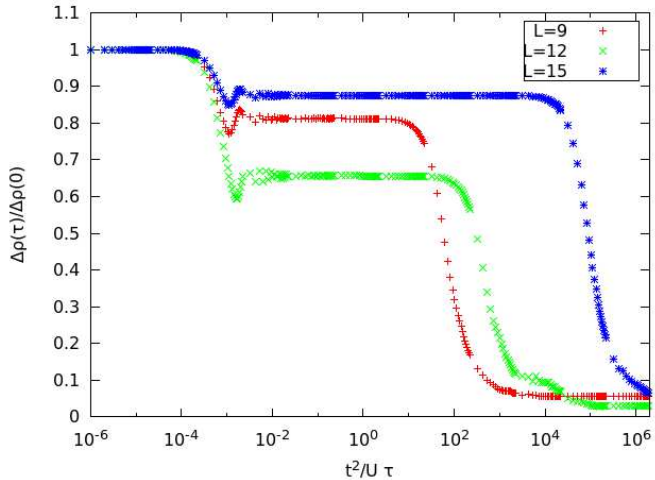}
\caption{Time evolution of the inhomogeneity for configurations containing first order resonances. The samples of length $L=9,15$ have only one particle  involved in the resonance; for $L=12$ two particles are involved. The presence of local resonances leads to partial relaxation processes at short time scales $\tau\approx O(t^{-1})$. Moreover, by comparing the plot with the inset of Fig.~\ref{Fig:Drinfinity}, one sees that the global relaxation times $\sim t^{-1}_{\rm eff}$ are reduced by a factor of $t/U$ per particle involved in resonances. As resonances are rare, this effect does not alter the fact that $t^{-1}_{\rm eff}$ diverges exponentially in the thermodynamic limit.}
\label{Fig::Inhomogeneity_resonances}
\end{figure}

Apart from increasing the effective hopping of the system, resonances result also in fast, partial relaxation processes through admixture. This diminishes the inhomogeneity plateau in $\langle \Delta \rho^2\rangle$ by an amount proportional to the density $\sim\rho$ of resonating configurations. This  effect is seen in Fig.~\ref{Fig::Inhomogeneity_resonances}, where the evolution of the inhomogeneity is plotted for configurations which include a resonance at first order in $t$.

Let us now determine the exponent $\alpha$ to leading order in the density $\rho\ll 1$, within perturbation theory. The simplest type of resonance is a pair of two consecutive intervals with lengths
\begin{equation}
\left(l,l+1\right)\textrm{ or }\left(l+1,l\right),
\end{equation}
as shown in Fig.~\ref{Fig::Resonance}. The probability of finding an interval of length $l$ in a random configuration of density $\rho$ is
\begin{equation}
P\left(l\right)=\rho\left(1-\rho\right)^{l-1}.
\end{equation}
There are 
\begin{equation}
N_{\textrm{1res}}=  2N \rho^{2}\sum_{l=1}^{\infty}\left(1-\rho\right)^{2l-1} +O(\rho^2) %=2N\rho\frac{1-\rho}{2-\rho}
=\rho N+O(\rho^{2})
\end{equation}
such resonances in a typical configuration $C$, where we neglect corrections due to overlapping pairs.
The factor of $2$ accounts for both possibilities $\left(l,l+1\right)$
and $\left(l+1,l\right)$. As discussed above, local configurations like this hybridize at first order in perturbation theory. Accordingly they reduce the power of $t$ in the effective tunneling by one each, which  yields
\begin{equation}
(\Delta\alpha)_{\rm 1res} =-\rho+O(\rho^{2}).
\end{equation}

The dominant reduction of $\alpha$ is, however, due to sequences of interval lengths of the form
\begin{equation}
\left(l,p_{1},...,p_{m},l+1\right),\label{sequence}
\end{equation}
where the $p_{i =1,...,m} \notin \{l-1,l,l+1\}$ are non-resonant with $l$ or $l+1$. If $m>1$, such configurations do not lead to strong hybridizations though, and thus they do not contribute significantly to the fast relaxation of the density inhomogeneity, $\Delta \rho^2$, which occurs before the long-time plateau. Nevertheless, they increase the effective hopping by introducing a small denominator
 in perturbation theory. Such a denominator is generically of order $t^2$, due to self-energies that arise in second order of perturbation theory. As discussed above, those typically lift the degeneracy present at the classical level. (For further discussion of higher order degeneracies, see Ref.~\cite{Us}). If two separated pairs of $l, l+1$ and $l', l'+1$ are interlaced, only one of them can be used to create a small denominator, however. The maximal number of resonances encountered in perturbation theory will usually be obtained by retaining the shorter of the two pairs. 

Let us now estimate the total number of resonant pairs of the form~(\ref{sequence}), which are not interlaced by shorter resonances.
To leading order the probability of finding such a sequence formed by $m+2$ intervals can be estimated  as $\rho$, multiplied by
the probability that there are no resonant sequences of shorter length
which interlace it. To compute this probability, we first impose the requirement that
the interval of length $l+1$ is not in resonance with the $m$ intervals
that follow it, which yields a factor $\left(1-\nicefrac{\rho}{2}\right)^{m}$.
Next we impose the requirement that the interval $p_{m}$ is not in resonance with
either $l+1$ nor with any of the subsequent $m-1$ intervals, which yields another
factor $(1-\nicefrac{\rho}{2})^{m}$. The preceding interval $p_{m-1}$ can be in resonance
with the interval $p_{m}$ (since such a resonance would be nested inside
the considered one) but not with $l+1$ or the following $m-2$ intervals.
This yields a factor $\left(1-\nicefrac{\rho}{2}\right)^{m-1}$. We iterate this procedure
up to interval $p_{1}$, and then square the resulting probability
since the same conditions apply on the left of the
sequence, too. This leads to
\bea
N_{2res}&\approx& N\rho\left(1-\frac{\rho}{2}\right)^{2m}\prod_{j=1}^{m}\left(1-\frac{\rho}{2}\right)^{2j}\nonumber\\
&=&N\rho\left(1-\frac{\rho}{2}\right)^{m^{2}+3m}\simeq N\rho e^{-\frac{\rho}{2}\left(m^{2}+3m\right)}.
\eea

The corresponding reduction in the exponent $\alpha$ can be estimated by summing the above over $m$
and approximating the sum as an integral:

\begin{equation}
(\Delta \alpha)_{2res} \simeq -2\rho\int_{1}^{\infty}dme^{-\frac{\rho}{2}\left(m^{2}+3m\right)}=-\sqrt{2\pi\rho}+O\left(\rho\right),
\end{equation}
where the factor of $2$ is due to the fact that each resonance typically increases the effective hopping by a factor $O\left(t^{-2}\right)$. 
This yields the dominant reduction of the tunneling exponent, $\alpha=1-\sqrt{2\pi\rho}$.

Note that the effective hopping could be computed by moving all particles either to the left or to the right. One might thus worry that the above result depends on this choice. However, one can check that in either construction the maximal number of small denominators encountered in calculating the perturbative matrix element is the same.

\section{\label{sec:bubbles}Effect of rare ergodic regions}

In recent works~\cite{Heidelberg2,Heidelberg3} it has been conjectured that in the thermodynamic limit delocalization might occur at any value of the hopping $t$, due to {\em non-perturbative} rare events within the configurations $C$. We briefly reproduce the argument below and discuss its potential relevance for the effects we have discussed above.

The argument starts from the observation that a random initial state will contain large, but very rare, regions where the particle and energy density are so low that a bulk system with the same parameters would be delocalized and ergodic. One then diagonalizes the Hamiltonian within such a bubble (considering it decoupled from the outside) to obtain effectively ergodic internal states. Further, one estimates the matrix element to displace the bubble by one site, at second order in the coupling to its neighbors. By making the volume of the bubble sufficiently large, the relevant energy denominators for such complex transitions become exponentially small in the volume of the relevant energy slice of the Hilbert space of the bubble. At the same time, the associated matrix elements decrease only with the square root of that volume. This suggests that the lateral displacement of a bubble is potentially a resonant process. Thus, big enough bubbles might eventually delocalize and form a mobile bath (i.e., an energy reservoir) for any other transition in the system. If this indeed happens, this effect would restore finite, even though very strongly suppressed, transport.

The above argument is not a proof of delocalization though, since it is very hard to control the effect of all the (much stronger) matrix elements which tend to diffuse the bubble and increase its  energy and particle density to a level where localization starts setting in. Whether such a bubble can dynamically evolve back to its initial shape and propagate resonantly from there, as assumed in the argument, or whether it becomes dynamically localized due to the coupling to many other environmental degrees of freedom, as in spin-bath problems~\cite{LeggettFisher, Stamp}, remains an open question. It is interesting to note, however, that, if such bubbles indeed do re-instate transport in disorder-free systems, analogous considerations to those above would rule out the many-body localization transitions at finite temperature, which were predicted in Ref.~\cite{BAA} for disordered systems. There delocalization might come about by the motion of rare hot and nearly ergodic bubbles which always exist in typical low-temperature states~\cite{Conserved3}.

In order to clarify the relevance of our predictions for experimental systems, we have estimated (see Appendix~\ref{sec:appbubbles}) the density $n_B$ of such rare bubbles for $t\ll t_c$ as
\begin{equation}
n_{B}\lesssim\exp\left\{ -2\left(\frac{t_{c}}{t}\right)^{\frac{1}{\beta}}\frac{\ln\frac{U}{t}}{\ln\left[\frac{1}{\rho}\left(\frac{t_{c}}{t}\right)^{\frac{1}{\beta}}\right]+1}\right\}.\label{PB}
\end{equation}
This tends to zero very rapidly as $t\rightarrow 0$. For $\rho=0.1$ and $t=0.01$ we find $n_B\lesssim 3\times10^{-4}$.
This shows that, deep enough in the localized phase, such effects can safely be neglected for realistic system sizes.

\section{\label{sec:conclusions}Experimental realizations and conclusions}

The simplest experimental realizations in which to observe the phenomenology described here, are strongly interacting cold atomic gases in one-dimensional optical lattices~\cite{Gavish04,Gadway10} or highly anisotropic spin chains and ladders, whose localization properties could be probed via hole burning techniques~\cite{Aeppli}.

While our calculation assumed periodic boundary conditions, the essence of interaction-induced localization will also be present in dense but randomly distributed cold atoms in a confining trap, which prevents the escape of particles at the boundaries. 
In this situation, we predict that the center of mass of an atomic cloud will respond to a tilt of the trap exponentially weakly, as it is governed by an effective hopping $t_{\rm eff}$ which is exponentially small in the article number.
%Similar systems as Eq.~(\ref{H})%, such as the one studied in Ref.~\cite{Us}, 
%can be realized .  
%The same physics could be realized in higher dimensions, even though localization is harder to achieve. 

In conclusion, we have shown that relaxation in an interacting quantum system without disorder can be exponentially slow in the system size, suggesting that in the thermodynamic limit the dynamics become genuinely non-ergodic. 
For the  power law interactions considered here, the ensuing quantum glass phase persists up to hopping strengths of the order of typical interaction energies between individual particles. The fluctuations of the latter tend to increase with thermal disorder. Therefore, temperature has a localizing tendency, in stark contrast to its dephasing role in disorder-dominated localization.

{\em Note added:} Recently we became aware of a related study~\cite{Norm} which finds an exponentially growing time scale for relaxation, in agreement with our results. The authors further report a scale-dependent relaxation time. We conjecture that the latter is a specific property of linear response, which is absent in our relaxation dynamics from random initial conditions and the dynamics studied in Ref.~\cite{Becca-Fabrizio}. Both are concerned with  strongly non-linear perturbations with respect to a homogeneous state.

\section*{Acknowledgements}

This research was supported in part by the National Science Foundation under Grant No. NSF PHY11-25915. M.M. and M.S. acknowledge the hospitality of KITP Santa Barbara and of the University of Basel, where part of this research was carried out.

\appendix
\section{\label{sec:appdecay}Temporal decay of spatial inhomogeneity}

We characterize the spatial inhomogeneity of the system by 
\begin{equation}
\Delta\rho_{\psi}^{2}\left(\tau\right)\equiv\frac{1}{L}\sum_{j=1}^{L}\left[\left\langle \psi\left(\tau\right)\right|\left(n_{j+1}-n_{j}\right)\left|\psi\left(\tau\right)\right\rangle \right]^{2},
\end{equation}
where $\left|\psi(\tau)\right\rangle = \exp[-iH\tau] \left |C\right\rangle$ is the state time evolved from the classical initial configuration $C$. For small hopping $t$, if we restrict $C$ to configurations without resonances, the only eigenstates with significant overlap with $C$ are the states in the miniband described by Eq.~(\ref{Miniband_state}).  Expanding in those eigenstates, labeled by $n,m$, we obtain
\bea
\Delta\rho^{2}\left(\tau\right)&=&\frac{1}{L}\sum_{j=1}^{L}\left[\sum_{n,m}e^{i\left(\varepsilon_{n}-\varepsilon_{m}\right)\tau}\left\langle C\left|n\right.\right\rangle \left\langle m\left|C\right.\right\rangle \left\langle n\right|\Delta\rho_{j}\left|m\right\rangle \right]^{2}\nonumber\\
&=&\frac{1}{L}\sum_{j=1}^{L}\sum_{n,m}\sum_{n^{\prime},m^{\prime}}e^{i\left[\left(\varepsilon_{n}+\varepsilon_{n^{\prime}}\right)-\left(\varepsilon_{m}+\varepsilon_{m^{\prime}}\right)\right]\tau}\left\langle C\left|n\right.\right\rangle \nonumber\\
&\times&\left\langle m\left|C\right.\right\rangle \left\langle n\right|\Delta\rho_{j}\left|m\right\rangle\\
&\times& \left\langle C\left|n^{\prime}\right.\right\rangle \left\langle m^{\prime}\left|C\right.\right\rangle \left\langle n^{\prime}\right|\Delta\rho_{j}\left|m^{\prime}\right\rangle,\nonumber
\eea
where the energies $\varepsilon_n$ are given by Eq.~(\ref{Miniband_energy}), and the overlaps with the initial configuration are given by
\begin{equation}
\left\langle C\left|m\right.\right> =\frac{1}{\sqrt{L}}.
\end{equation}

Since the operators $n_i$ are diagonal in the basis of classical configurations, the matrix elements of the site occupations are
\bea
\left\langle n\right|n_{j}\left|m\right\rangle& =&\frac{1}{L}\sum_{k,k^{\prime}=0}^{L-1}e^{i\frac{2\pi}{L}\left(m+\phi\right)k}e^{-i\frac{2\pi}{L}\left(n+\phi\right)k^{\prime}}\nonumber\\
&\times&\left\langle C\right|T^{-k^{\prime}}n_{j}T^{k}\left|C\right\rangle
\\ &=&\frac{1}{L}\sum_{k=0}^{L-1}e^{i\frac{2\pi}{L}\left(m-n\right)k}\left\langle C\right|n_{j+k}\left|C\right\rangle ,\nonumber
\eea
where $T$ is the translation operator. Then the expression for the inhomogeneity becomes
\begin{eqnarray}
\Delta\rho_{\psi}^{2}\left(T\right) & = & \frac{1}{L^{5}}\sum_{j=1}^{L}\sum_{m,n=0}^{L-1}\sum_{n^{\prime},m^{\prime}=0}^{L-1}e^{i\left[\left(\varepsilon_{n}+\varepsilon_{n^{\prime}}\right)-\left(\varepsilon_{m}+\varepsilon_{m^{\prime}}\right)\right]\tau}\nonumber\\
 & &\times  \sum_{k=0}^{L-1}\sum_{k^{\prime}=0}^{L-1}e^{i\frac{2\pi}{L}\left(m-n\right)k}e^{i\frac{2\pi}{L}\left(n^{\prime}-m^{\prime}\right)k^{\prime}}\\
&\times&\left\langle C\right|\left(n_{j+k+1}-n_{j+k}\right)\left|C\right\rangle \nonumber\\
&\times&\left\langle C\right|\left(n_{j+k^{\prime}+1}-n_{j+k^{\prime}}\right)\left|C\right\rangle.\nonumber
\end{eqnarray}

This leads to Eq.~(\ref{DrhoT}) of the main text, upon using the density auto-correlation function,
\be
G\left(k-k^{\prime}\right)\equiv\frac{1}{L}\sum_{j}\left\langle C\right|n_{j+k}\left|C\right\rangle \left\langle C\right|n_{j+k^{\prime}}\left|C\right\rangle,
\ee 
in the initial state.

\section{\label{sec:appbubbles}Density of rare, nearly ergodic bubbles}

We consider an initial random state which includes an ``ergodic bubble" where the local energy density is below the critical threshold for bulk localization (see Fig.~\ref{Fig::MBLDiagram}). We assume the global density of particles $\rho$ to be small, and  $t$ sufficiently smaller than the delocalization threshold $t_c\left(\rho\right)$, as estimated in Eq.~(\ref{t_C}) for states with roughly homogeneous density distributions.  Recalling that %$t_{c}\propto\rho^{\beta+1}$ 
$t_{c}\propto\rho^{\beta}$, the density $\rho_B$ in the ergodic bubble should be smaller than
\begin{equation}
\label{rhoB}
\frac{\rho_{B}}{\rho}\lesssim\left(\frac{t}{t_c}\right)^{\frac{1}{\beta}}.
\end{equation}

Denoting by $L_B$ the number of sites in the bubble, the dimension of the Hilbert space $\mathcal{H}_{B}$ of  internal states with $\rho_BL_B$ particles is
\bea
{\rm dim}(\mathcal{H}_{B})&=&\begin{pmatrix}L_{B}\\
\rho_{B}L_{B}
\end{pmatrix}\approx \exp[\rho_B(1-\ln(\rho_B)) L_B]
\equiv\kappa^{L_{B}},\nonumber\\&& \rho_B\ll 1.
\eea
Since we assume $\rho_B<\rho$ to be very small, $\kappa$ is very close to $1$, such that the phase space of such bubbles grows slowly with their size. Consequently, very large regions are necessary to obtain small enough level spacings that might potentially induce delocalization of the bubble.

The minimal size $L_B$ is estimated from the hybridization between an initial  bubble state $\psi_i$ and a final state $\psi_f$ in which the bubble has moved by one site.
Delocalization may potentially occur if the admixture of $\psi_f$ to $\psi_i$ is large in second order in perturbation theory, i.e., if
\begin{equation}
\sum_{\psi_B} t^{2}\frac{\left\langle \psi_{f}\right|O \left|\psi_{B}\right\rangle \left\langle \psi_{B}\right|O\left|\psi_{i}\right\rangle }{\left(E_{i}-E_{B}\right)\left(E_{B}-E_{f}\right)}\gtrsim1\label{mat_el},
\end{equation}
where $\left|\psi_{B}\right\rangle$ runs over intermediate states, and $t\times O$ is the part of the hopping Hamiltonian that couples the bubble to the surrounding degrees of freedom. Let us first estimate the matrix elements of the hopping: making the generous assumption that the bubble is internally fully ergodic and that its eigenstates satisfy the eigenstate thermalization hypothesis~\cite{ETH}, matrix elements with a generic local operator can be argued to scale as 
\begin{equation}
\left\langle \phi\right|O\left|\chi\right\rangle \sim\frac{1}{\sqrt{\textrm{dim}\left(\mathcal{H}_{B}\right)}}\sim\kappa^{-\frac{L_{B}}{2}},
\end{equation}
where $\phi,\chi$ label generic internal eigenstates. In order to minimize the energy denominators in~(\ref{mat_el}), one should optimize the intermediate and final states, which yields
\begin{equation}
{\rm min}_{\chi} |E_{\chi}-E_{\phi}| \sim\frac{U}{\textrm{dim}\left(\mathcal{H}_{B}\right)}\sim U\kappa^{-L_{B}}.
\end{equation}

Inserting these estimates into Eq.~(\ref{mat_el}), we obtain a condition on $L_B$:

\begin{equation}
L_{B}\gtrsim 2\frac{\log\left({U}/t\right)}{\log\kappa}=\frac{2}{\rho_B}\frac{\log\left({U}/t\right)}{\log(1/\rho_{B})+1}.
\end{equation}
Note that the required length diverges logarithmically in the limit $t\to 0$, implying that these bubbles are nonperturbative in nature. In this aspect they bear some resemblance to rare regions in Griffiths phases. 

The density $n_B$ of such large bubbles  is given by the probability of finding only $\rho_B L_B$ particles in a region of length $L_B$, while the global density is $\rho$. For small $\rho$ and $t$ this is given by
\bea
n_{B}&\approx&\begin{pmatrix}L_{B}\\
\rho_{B}L_{B}
\end{pmatrix}\rho^{\rho_{B}L_{B}}\left(1-\rho\right)^{L_{B}\left(1-\rho_{B}\right)}\\&\approx&\exp\left[-L_{B}\left(\rho-\rho_{B}-\rho_{B}\log\frac{\rho}{\rho_{B}}\right)\right].\nonumber
\eea

In the regime $t\ll t_c$ (and thus $\rho_B\ll \rho$) this can be approximated as $n_B\approx \exp(- \rho L_B)$.
Using the bound on $\rho_B$ from Eq.~(\ref{rhoB}) we find an upper bound on the density of ergodic bubbles,
\begin{equation}
n_{B}\lesssim\exp[-\rho L_{B}]\lesssim\exp\left\{ -2\left(\frac{t_{c}}{t}\right)^{\frac{1}{\beta}}\frac{\log\frac{U}{t}}{\log\left[\frac{1}{\rho}\left(\frac{t_{c}}{t}\right)^{\frac{1}{\beta}}\right]+1}\right\} ,
\end{equation}
which is the expression given in Eq.~(\ref{PB}).
This is exponentially small and non-perturbative in the limit $t\to 0$.
For system sizes $L\ll 1/n_B$, such effects are irrelevant, since a typical realization will not contain any such bubbles.

\end{document}